\documentclass[10pt,a4paper,english,aps]{revtex4}
\usepackage[T1]{fontenc}
\usepackage[latin1]{inputenc}
\usepackage{babel}
\usepackage{graphics}

\makeatletter

\providecommand{\LyX}{L\kern-.1667em\lower.25em\hbox{Y}\kern-.125emX\@}

\usepackage[T1]{fontenc}
\usepackage[latin1]{inputenc}
\usepackage{babel}
\usepackage{graphics}

\makeatletter

\usepackage[T1]{fontenc}
\usepackage[latin1]{inputenc}
\usepackage{babel}
\usepackage{graphics}



\makeatother

\makeatother
\begin{document}

\title{Tracking and coupled dark energy as seen by WMAP}

\author{Luca Amendola\( ^{1} \) \& Claudia Quercellini \( ^{1,2} \)}

\affiliation{\( ^{1} \)INAF - Osservatorio Astronomico di Roma, \\
Via Frascati 33, 00040 Monte Porzio Catone - Italia\\
\( ^{2} \)Dip. di Fisica, Universita' di Roma Tor Vergata, Via della
Ricerca Scientifica 1, 00133 Roma - Italia}

\begin{abstract}
The satellite experiment WMAP has produced for the first time a high-coverage,
high-resolution survey of the microwave sky, releasing publicly available
data that are likely to remain unrivalled for years to come. Here
we compare the WMAP temperature power spectrum, along with an exhaustive
compilation of previous experiments, to models of dark energy that
allow for a tracking epoch at the present, deriving updated bounds
on the dark energy equation of state and the other cosmological parameters.
Moreover, we complement the analysis by including a coupling of the
dark energy to dark matter. The main results are: \emph{a}) the WMAP
data alone constrain the equation of state of tracking dark energy
to be \( w_{\phi }<-0.67(-0.49) \) to 68\%(95\%) (confining the analysis
to \( w_{\phi }>-1 \)), which implies for an inverse power-law potential
an exponent \( \alpha <0.99(2.08) \); \emph{b}) the dimensionless
coupling to dark matter is \( |\beta |<0.07(0.13) \). Including the
results from the supernovae Ia further constrains the dark energy
equation of state.
\end{abstract}
\maketitle

\section{Introduction}

The WMAP satellite has just released the first full sky maps produced
after more than one year of operation (Bennett et al. 2003). These
data mark an important step in cosmology: they represent in fact the
first publicly available multi-band high-resolution full-sky survey
of the microwave sky. The WMAP survey is likely to remain the highest-quality
CMB temperature anisotropy survey until the launch of the Planck satellite
not earlier than 2007. Although the exercise of parameter fitting
with each new experiment has produced in the last few years a copious
literature, the new WMAP data are certainly on a class of its own
and deserve to be taken fully into account. Here we address the question
of what the first year of WMAP data can tell us about the property
of dark energy.

Five years after the first observational hints about the existence
of a dominant component of unclustered matter with negative pressure
(Riess et al. 1998, Perlmutter et al. 1999), the so-called dark energy
or quintessence (Wetterich 1988; Ratra \& Peebles 1988; Frieman et
al. 1995; Ferreira \& Joyce 1997; Caldwell et al. 1998), there are
still very few indications as to its nature. The main reason, perhaps,
is that we lack any specific theoretical suggestion on the properties
of the dark energy, i.e. on its self-interaction potential and on
how it interacts with the other cosmological components. Several works
have tried to constrain dark energy models with pre-WMAP high resolution
CMB data, leading to bounds on the dark energy equation of state.
Unless strong priors on the other cosmological parameters are imposed,
however, these bounds turned out to be rather weak. In particular,
the Hubble constant is strongly degenerated with the dark energy present
equation of state, \( w_{\phi } \) (Huey et al. 1999). For instance,
with a flat prior \( h\in (0.45-0.9) \), Amendola et al. (2002) found
that essentially all values \( w_{\phi }<0 \) were allowed, while
\( w_{\phi }=-0.65\pm 0.2 \) if \( h=0.65\pm 0.05 \). Similarly,
adopting the HST value \( h=0.72\pm 0.08 \) (Freedman et al. 2001)
and using updated data Melchiorri et al. (2002) found \( w_{\phi }\in (-0.45,-1.81) \)
at 95\% c.l. (they included also values \( w_{\phi }<-1 \)). Other
works that addressed the same issue includes Corasaniti \& Copeland
(2002), Baccigalupi et al. (2002). With pre-WMAP data, only the imposition
of a low value of \( h \) (e.g. \( h<0.7 \)) could allow to distinguish
dynamical dark energy from a pure cosmological constant (or, more
precisely, from a fluid possessing an equation of state close to \( w_{\phi }=-1 \)
at the present). 

In reality, all the constraints for CMB presented in the literature
on the dark energy equation of state are necessarily somewhat model-dependent,
since the fundamental quantity, namely the angular-diameter distance
to last scattering, in general depends on \( w_{\phi } \) through
two integrals over the cosmic evolution that do not allow to distinguish
among different parameters along degeneracy curves. Although at low
\( \ell  \)s the geometric degeneracy could in principle be broken
by the ISW effect, the cosmic variance and the weakness of the effect
make this possibility unrealistic in most cases. Here, as in several
other works on the topics, we restrict therefore our attention to
two classes of models which are at the same time general (in the sense
of covering most modelizations) and simple. The initial conditions
are choosen so that the trajectories reach the tracking solutions
(notice that off-tracking solutions are indistinguishable from a cosmological
constant). First, we consider dark energy as a scalar field with inverse
power-law potentials \( V\sim \phi ^{-\alpha } \) (Ratra \& Peebles
1988), which recovers the exponential potentials for large values
of \( \alpha  \) ; in this case, the equation of state at the present
is roughly constant\begin{equation}
\label{wphi}
w_{\phi }=\frac{p_{\phi }}{\rho _{\phi }}\approx -\frac{2}{\alpha +2}
\end{equation}
due to the tracking mechanism (Steinhardt et al. 1999). This approximates
also the cases in which the dark energy is a perfect fluid with constant
\( w_{\phi }, \) since the angular size of the acoustic horizon depend
on the equation of state only, and includes also the cases in which
\( w_{\phi } \) is the \( \Omega _{\phi } \)-weighted average of
a slowly-varying function \( \hat{w}_{\phi }(a) \) (see e.g. Doran
et al. 2001). 

The WMAP team analysed their power spectrum including dark energy
with a constant equation of state (and no coupling). They found for
the equation of state \( w_{\phi }<-0.5 \) with CMB data alone and
\( w_{\phi }<-0.78 \) including supernovae (Spergel et al. 2003).
Although dark energy with an inverse power-law potential has an equation
of state which is constant only near the present epoch, we will recover
very similar results for this class of models, as expected.

In the second class we include models of dark energy coupled to dark
matter (Amendola 2000; Amendola et al. 2002). This class of models,
in which we include the conformally related Brans-Dicke Lagrangians,
has been widely studied in the context of dark energy (Uzan 1999,
Chiba 1999, Chen \& Kamionkowski 1999, Baccigalupi et al. 2000, Holden
\& Wands 2000, Chimento et al. 2000, Billyard \& Coley 2000, Bean
\& Magueijo 2000, Esposito-Farese \& Polarsky 2001, Sen \& Sen 2001,
Gasperini et al. 2002, Pietroni 2002) It is important to stress that
the behavior of coupled dark energy cannot be modeled simply by some
choice of the equation of state for \( \phi  \) because the interaction
induces on dark matter an effective equation of state different from
zero. In this case dark energy mediates a long-range scalar interaction
that modifies the gravity felt by dark matter particles through a
Yukawa-type term (Damour et al. 1990). On Newtonian scales, the interaction
simply renormalizes Newton's constant for dark matter \begin{equation}
\label{yuk}
G'=G(1+\frac{4\beta ^{2}}{3})
\end{equation}
where \( \beta  \) is the dimensionless coupling, while the baryons
remain uncoupled (or very weakly coupled, as required by experimental
constraints, see e.g. Hagiwara et al. 2002). Then, dark energy effectively
violates the equivalence principle, but in a way that is locally unobservable.
For as concerns us, the main effect of the coupling is to induce an
exchange of energy between the two dominating dark components so that,
after equivalence but before acceleration, the universe enters a regime
in which not only the equation of state (as in the tracking case)
but also the density parameters \( \Omega _{m},\Omega _{\phi } \)
are constant. In Amendola (2001) this regime was called \( \phi  \)MDE,
to remark the fact that matter is not the only dominating component.
This epoch ends when the dark energy enters the tracking regime, the
potential energy takes over and accelerates the expansion. During
\( \phi  \)MDE we have \( \Omega _{\phi }=4\beta ^{2}/3 \) and an
effective equation of state for the coupled fluid dark matter/dark
energy \begin{equation}
\label{we}
w_{e}=\frac{p_{tot}}{\rho _{tot}}=\frac{4\beta ^{2}}{9}
\end{equation}
 Note that this behavior is an example of {}``early quintessence{}''
(Caldwell et al. 2003), in which the dark energy density is not negligible
at last scattering. Scope of this paper is to constrain both equations
of state, the present one, \( w_{\phi } \) and the past one, \( w_{e} \).
This will put limits on the two {}``fundamental{}'' parameters \( \alpha ,\, \beta  \).
A value \( \alpha \not =0 \) would imply that the dark energy is
not a pure cosmological constant, while a value \( \beta \not =0 \)
would imply a large-scale violation of the equivalence principle.
In the following we will use interchangeably the parameters \( w_{\phi },w_{e} \)
or \( \alpha ,\, \beta  \). In Fig. 1 we compare the background trajectories
for the uncoupled (\( \beta =0 \)) and coupled (\( \beta =0.1 \))
cases, assuming \( \alpha =1,\omega _{b}=0.02,\omega _{c}=0.1,h=0.7 \).
Notice the plateau of constant \( \Omega _{\phi } \) in the coupled
model.

We start with assuming no coupling, \( \beta =0 \), and varying \( w_{\phi } \)
along with \( h,\omega _{c}=\Omega _{c}h^{2},\omega _{b}=\Omega _{b}h^{2} \)
and the slope \( n_{s} \) of the primordial fluctuations. Then we
include \( \beta  \) and constrain both \( w_{\phi } \) and \( w_{e} \).
We assume a flat space throughout. Notice that our \( \Omega _{c} \)
refers to cold dark matter only, so that the total matter content
is \( \Omega _{m}=\Omega _{b}+\Omega _{c} \).
\begin{figure}[t]
{\centering \resizebox*{!}{12cm}{\includegraphics{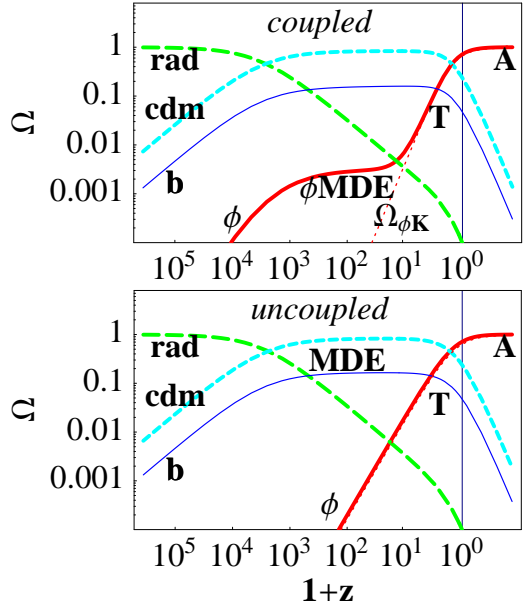}} \par}

\caption{Behavior of the energy densities for the coupled (\protect\( \beta =0.1\protect \),
top) and uncoupled (\protect\( \beta =0\protect \), bottom) model.
Notice, in the coupled case, the \protect\( \phi \protect \)MDE in
which \protect\( \Omega _{\phi }\protect \) is almost constant and
dominated by the kinetic energy of the scalar field. The label \protect\( T\protect \)
denotes the tracking behavior, the label \protect\( A\protect \)
denotes the final future attractor. }
\end{figure}

\section{Data and method}

The class of models considered in this paper is the same as in Amendola
et al. (2002). We compare the models to the combined power spectrum
estimated by WMAP (Hinshaw et al. 2003). To derive the likelihood
we adopt a version of the routine described in Verde et al. (2003),
which takes into account all the relevant experimental properties
(calibration, beam uncertainties, window functions, etc). Since the
likelihood routine employs approximations that work only for spectra
not too far from the data, we run it only for models whose \( \chi ^{2} \)
is less than four times the degrees of freedom. We experimented with
increasing the \( \chi ^{2} \) cut and found no important variations.

We also compare the data to the whole set of pre-WMAP data. To do
this, we use the power spectrum provided by Wang, Tegmark and Zaldarriaga
(2002), which is a compression of essentially all the available data
up to mid-2002 in 25 \( \ell  \)-bins from \( \ell =2 \) to \( \ell =1700 \),
complete of correlation matrix and window functions (calibration and
beam uncertainty are included in the correlation matrix). The main
entries in this compilation are COBE (Bennett et al. 1996), Boomerang
(Netterfield et al. 2002), DASI (Halverson et al. 2002), Maxima (Lee
et al. 2002), CBI (Pearson et al. 2002), VSA (Scott et al. 2002).
To this we add Archeops (Benoit et al. 2002) with its correlation
matrix, window functions, beam and calibration uncertainties. For
the pre-WMAP data we assumed no reionization (\( \tau =0 \)) because
this was the best fit before WMAP.

For the pre-WMAP data we integrated out analytically the calibration
and beam uncertainties and the overall normalization. Let us denote
with \( N_{e},\sigma _{e},\sigma _{w,e} \) the correlation matrix,
the calibration uncertainty and the beam uncertainty, respectively,
of the \( e \)-th experiment, and with \( I_{e}=\{1,1,1,...,\} \)
a vector of dimension equal to the number of bins in the \( e \)-th
experiment. Moreover, let \( C_{\ell ,t}\prime  \) be the theoretical
CMB spectrum binned in the \( \ell  \)-th bin with the experimental
window function. Then one gets the remarkably simple likelihood function
(see Appendix)\begin{equation}
\label{ourlik}
L=\exp \left[ -\frac{1}{2}\left( \gamma -\frac{\beta ^{2}}{\alpha }\right) \right] 
\end{equation}
where\begin{eqnarray}
\alpha  & = & \sum _{e}I_{e}^{T}M_{e}^{-1}I_{e}\\
\beta  & = & \sum _{e}I_{e}^{T}M_{e}^{-1}(Z_{t}-Z_{d})\label{abg} \\
\gamma  & = & \sum _{e}(Z_{t}-Z_{d})^{T}M_{e}^{-1}(Z_{t}-Z_{d})
\end{eqnarray}
 where the matrix \( M_{e} \) is\begin{equation}
\label{gencorr}
M_{e}=(C_{\ell ,d}^{T}C_{\ell ,d})^{-1}N_{e}+\sigma _{e}^{2}I_{e}^{T}I_{e}+\sigma _{w,e}A^{T}A
\end{equation}
and where \( Z_{t}=\log (C_{\ell ,t}\prime ) \) and \( Z_{d}=\log C_{\ell ,d} \)
and where the vector \( A \) expresses the dependence on \( \ell  \)
of the beam uncertainty (in the case of a Gaussian beam, \( A=\{\ell _{1}^{2},\ell _{2}^{2},...,\ell _{i}^{2}\} \)).
This is the likelihood we use for the pre-WMAP data.

Our theoretical model depends on two scalar field parameters, four
cosmological parameters and the overall normalization: \begin{equation}
\alpha ,\beta ,n_{s},h,\omega _{b},\omega _{c},A.
\end{equation}
 As anticipated, to save computing time we found it necessary to restrict
the analysis to a flat space; moreover, we fixed the optical depth
to \( \tau =0.17 \), the best fit found by MAP (Spergel et al. 2003).
We derived the likelihood also for \( \tau =0.1 \) and found no significative
differences for as concerns the two dark energy parameters \( \alpha ,\, \beta  \).
The initial conditions for the scalar field are found iteratively
for each set of cosmological parameters. The overall normalization
has been integrated out numerically. Since \( \alpha ,\beta  \) determine
uniquely the present and past equations of state, we will present
the likelihood also as function of \( w_{\phi },w_{e} \) . We calculate
the theoretical \( C_{\ell ,t} \) spectra by a modified parallelized
CMBFAST (Seljak \& Zaldarriaga 1996) code that includes the full set
of coupled perturbation equations (see Amendola 2000 and Amendola
\& Tocchini-Valentini 2001). The other parameters are set as follows:
\( T_{cmb}=2.726K, \) \( Y_{He}=0.24,N_{\nu }=3.04 \). 

We evaluated the likelihood on two grids of roughly \( 50,000 \)
models each (for each normalization): a sparse grid that covers a
broad volume was used as a preliminary exploration; a second denser
grid centered on the peaks of the first was then used for the actual
calculations. For the first grid we adopted the following top-hat
broad priors: \( \beta \in (0,0.3), \) \( \quad \alpha \in (0.25,20),\quad  \)\ \( n_{s}\in (0.8,1.2), \)
\( \quad \omega _{b}\in (0.005,0.04),\quad  \)\ \( \omega _{c}\in (0.05,0.3) \)
. For the Hubble constant we adopted the top-hat prior \( h\in (0.5,0.9); \)
we also employed the HST result (Freedman et al. 2001) \( h=0.72\pm 0.08 \)
(Gaussian prior) but found only minor differences, since the WMAP
results are already very close to the HST distribution. The same age
constraint (\( >10 \) Gyr) used in most previous analyses is adopted
here.

\section{Uncoupled tracking dark energy}

We begin by putting \( \beta =0 \) and assuming \( V(\phi )=A\phi ^{-\alpha } \).
In Fig. 2 we show the likelihood for each parameter, marginalizing
in turn over the others. The dotted line is for the HST prior \( h=0.72\pm 0.08 \).
Notice that, due to the degeneracy between \( w_{\phi } \) and \( h \),
the limits on \( w_{\phi } \) are rather weak, especially if no prior
is assumed. The numerical results are in Table I. Here and in the
following the limits are always at the 68\%(95\%) c.l. while the errors
are at 68\% c.l.. As expected, we find results very close to those
in Spergel et al. (2003). The small residual differences are probably
due to the fact we use a grid instead of a Markov chain and fix \( \tau  \)
instead of marginalizing over it.

\(  \)

\(  \)

\( \qquad  \) \( \qquad  \) \( \qquad  \) \( \qquad  \) \( \qquad  \)
\( \qquad  \)\begin{tabular}{|c|c|c|}
\hline 
parameter&
WMAP&
WMAP+HST\\
\hline
\hline 
\( w_{\phi } \)&
\( <-0.68(-0.51) \)&
\( <-0.73(-0.55) \)\\
\hline 
\( \alpha  \)&
\( <0.94(1.92) \)&
\( <0.74(1.64) \)\\
\hline 
\( h \)&
0.69\( ^{+0.04}_{-0.05} \)&
0.70\( ^{+0.03}_{-0.05} \)\\
\hline 
\( n_{s} \)&
1.01\( \pm  \)0.022&
1.01\( \pm  \)0.022\\
\hline 
\( \omega _{b} \)&
0.0247\( \pm  \)0.001&
0.0247\( \pm  \)0.0008\\
\hline 
\( \omega _{c} \)&
0.12\( _{-0.01}^{+0.02} \)&
0.12\( _{-0.01}^{+0.015} \)\\
\hline 
\multicolumn{3}{|c|}{Table I. Uncoupled dark energy.}\\
\hline
\end{tabular}

\(  \)
\begin{figure}[t]
{\centering \resizebox*{20cm}{15cm}{\includegraphics{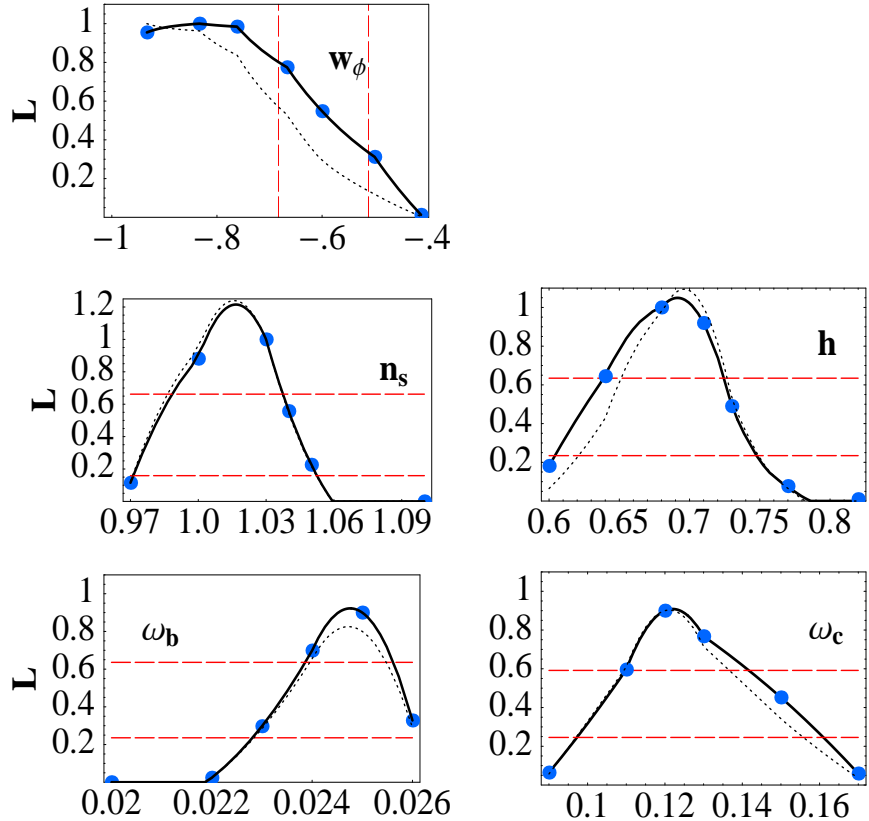}} \par}

\caption{Likelihood functions for uncoupled dark energy models, in arbitrary
units. In each panel the other parameters have been marginalized.
The dotted lines are for the HST prior on the Hubble constant. The
horizontal long-dashed lines are the confidence levels at 68\% and
95\%. The vertical long-dashed lines in the panel for \protect\( w_{\phi }\protect \)
mark the upper bounds at 68\% and 95\% confidence levels.}
\end{figure}

\section{Coupled dark energy}

In the flat conformal FRW metric \( ds^{2}=-dt^{2}+a^{2}dx_{i}dx^{i} \)
the scalar coupling modifies the conservation equations of dark matter
and dark energy as follows: \begin{eqnarray}
\ddot{\phi }+3H\dot{\phi }+a^{2}U_{,\phi } & = & C\rho _{c},\\
\dot{\rho }_{c}+3H\rho _{c} & = & -C\rho _{c}\dot{\phi },\label{kg} 
\end{eqnarray}
 where \( H=\dot{a}/a \). We assume that the baryons are not directly
coupled to the dark energy (otherwise local gravity experiment would
reveal a fifth force, see Damour et al. 1990; the radiation is automatically
uncoupled with this coupling, see e.g. Amendola 1999). The dimensionless
coupling \begin{equation}
\beta ^{2}=\frac{3C^{2}}{16\pi G},
\end{equation}
can be seen as the ratio of the dark energy-dark matter interaction
with respect to gravity (Damour \& Nordtvedt 1993, Wetterich 1995).
In Amendola (2001) we have shown that the dynamics of the system is
insensitive to the sign of \( \beta  \), since the \( \phi  \)MDE
and the tracking phases do not depend on it. We will consider only
\( \beta >0. \)

It is important to observe that the CMB bounds on \( \beta  \) depend
on the existence of the \( \phi  \)MDE. This epoch has several features
that distinguish it from tracking: it is very long (from equivalence
to \( z\approx 10 \)); it depends only on the kinetic energy of the
dark energy, and therefore is independent of the dark energy potential;
it cannot be avoided even assuming an extremely small initial dark
energy density (contrary to the tracking that may be avoided by selecting
initial conditions with very low scalar field energy -- the so-called
{}``undershooting{}'' trajectories ). Most importantly, as shown
in Amendola et al. (2002), suffers less from the geometrical degeneracy
that plague \( w_{\phi } \), since \( \Omega _{\phi } \) at decoupling
is non-zero (although there still is a degeneracy \( \beta -h \)
in the sense that larger \( h \) are compensated by larger \( \beta  \)).

In Fig. 3 we show the likelihood for \( w_{\phi },w_{e} \), marginalizing
over the other parameters. We find the following constraints at 95\%
c.l.:\begin{equation}
\alpha <2.08,\quad \beta <0.13.
\end{equation}
 Notice that this implies that an exponential potential (corresponding
to \( \alpha \to \infty  \)) is rejected even for \( \beta \not =0 \).
In place of \( \alpha  \) and \( \beta  \) we can use as well as
likelihood variables the equation of state during tracking and during
\( \phi  \)MDE, respectively, using Eqs. (\ref{wphi}) and (\ref{we}).
Then we obtain at the 95\% c.l.-\begin{equation}
\label{cons}
-1\leq w_{\phi (tracking)}<-0.49,\quad 0\leq w_{e(\phi MDE)}<0.0075.
\end{equation}
This shows that the effective equation of state during \( \phi  \)MDE,
i.e. between equivalence and tracking, is close to zero (as in a pure
matter dominated epoch) to less than 1\%. The striking difference
between the level of the two constraints in (\ref{cons}) indicates
that the CMB is more sensitive to the dark energy coupling than to
its potential, as emphasized in Amendola et al. (2003). In Fig. 4
we report the likelihood for all the parameters, contrasting the WMAP
estimation with the pre-WMAP one (the limit on \( \beta  \) stated
in Amendola et al. 2003 was slightly different because here we include
more pre-WMAP data). As it can be seen, WMAP puts quite stronger limits
on \( w_{\phi } \) and on \( \beta  \). In particular, the likelihood
for \( w_{\phi } \) now vanishes for \( w_{\phi }>-0.4 \).

The other parameters are given in Table II. It appears that the limits
on the cosmological parameters \( n_{s,}\omega _{b},\omega _{c} \)
are almost independent of \( \beta  \), while a non-zero \( \beta  \)
favors higher \( h \). The degeneracy \( \beta -h \) is reported
in Fig. 5.

\(  \)

\(  \)

\begin{figure*}[t]
{\centering \resizebox*{!}{10cm}{\includegraphics{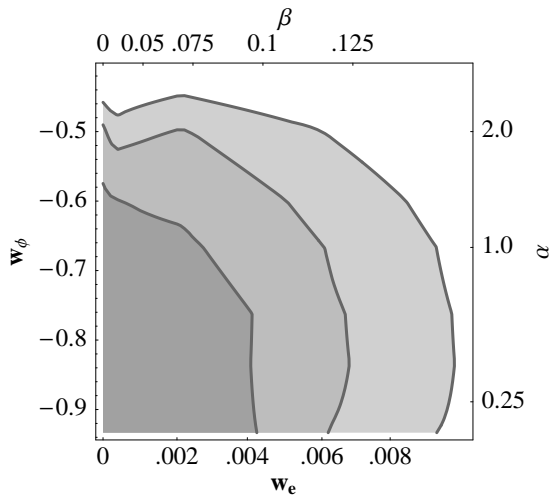}} \par}

\caption{Likelihood contour plots in the space \protect\protect\( w_{\phi (tracking)}\protect \protect \),\protect\protect\( w_{e(\phi MDE)}\protect \protect \)
(or \protect\protect\protect\protect\protect\( \alpha ,\beta \protect \protect \protect \protect \protect \))
marginalizing over the other parameters at the 68,95 and 99\% c.l..}
\end{figure*}

\begin{figure*}[t]
{\centering \resizebox*{20cm}{15cm}{\includegraphics{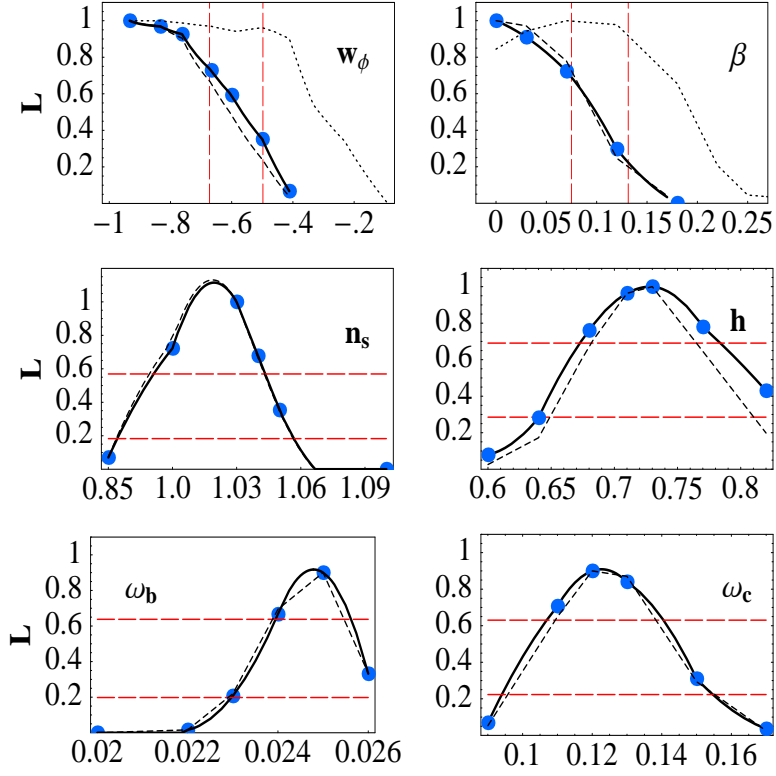}} \par}

\caption{Marginalized likelihood for tracking trajectories. The solid curves
are for the WMAP data, the short-dashed curves are for the HST prior,
and the dotted curves in the panels for \protect\( w_{\phi }\protect \)
and \protect\( \beta \protect \) for the pre-WMAP compilation . The
horizontal long-dashed lines are the confidence levels at 68\% and
95\%. The vertical long-dashed lines in the panel for \protect\( w_{\phi }\protect \)
mark the upper bounds at 68\% and 95\% confidence levels.}
\end{figure*}

\begin{figure}[t]
{\centering \resizebox*{!}{10cm}{\includegraphics{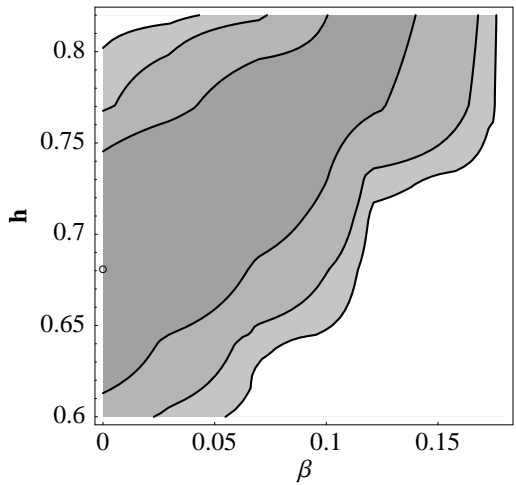}} \par}

\caption{Likelihood for \protect\( \beta ,\, h\protect \). This shows the
residual degeneracy between the two parameters due to the geometric
degeneracy in the angular diameter distance to last scattering.}
\end{figure}
\( \qquad  \) \( \qquad  \) \( \qquad  \) \( \qquad  \) \begin{tabular}{|c|c|c|c|}
\hline 
parameter&
WMAP&
WMAP+HST&
pre-WMAP\\
\hline
\hline 
\( w_{\phi } \)&
\( <-0.67(-0.49) \)&
\( <-0.69(-0.52) \)&
\( <-0.50(-0.25) \)\\
\hline 
\( \alpha  \)&
\( <0.99(2.08) \)&
\( <0.90(1.84) \)&
\( <2.0(6.0) \)\\
\hline 
\( w_{e} \)&
\( <0.0025(0.0075) \)&
\( <0.0023(0.0075) \)&
\( <0.0075(0.016) \)\\
\hline 
\( \beta  \)&
\( <0.075(0.13) \)&
\( <0.072(0.13) \)&
\( <0.13(0.19) \)\\
\hline 
\( h \)&
0.73\( \pm  \) 0.05&
0.73\( \pm  \) 0.04&
>0.62(0.55)\\
\hline 
\( n_{s} \)&
1.019\( \pm  \) 0.025&
1.018\( \pm 0.025 \)&
0.97\( \pm  \)0.03\\
\hline 
\( \omega _{b} \)&
0.0247\( \pm  \) 0.0008&
0.0250\( \pm  \)0.0008&
0.021\( \pm  \)0.003\\
\hline 
\( \omega _{c} \)&
0.123\( \pm  \) 0.016&
0.120\( \pm  \) 0.016&
0.12\( \pm  \)0.04\\
\hline 
\multicolumn{4}{|c|}{Table II. Coupled dark energy.}\\
\hline
\end{tabular}

\(  \)

\(  \)

In Fig. 6 we show the contour plot of the likelihood \( L(\Omega _{m},w_{\phi }) \),
where \( \Omega _{m}=\Omega _{c}+\Omega _{b} \), marginalizing over
the other parameters. We also add the confidence region from the Hubble
diagram of SNIa (we used the fit C of Perlmutter et al. 1999, plus
the supernova SN1997ff at \( z\approx 1.75 \) , Benitez et al. 2002).
The product of the two likelihood functions is shown in the same figure.
It turns out that \( \Omega _{\phi }=0.67\pm 0.05 \) and \( w_{\phi }<-0.76 \)
(95\%c.l.) (see Fig. 7). The limit on \( w_{\phi } \) is very close
to that obtained in Spergel et al. (2003): this shows that this bound
is almost independent of the coupling \( \beta . \)

\begin{figure}[t]
{\centering \resizebox*{!}{10cm}{\includegraphics{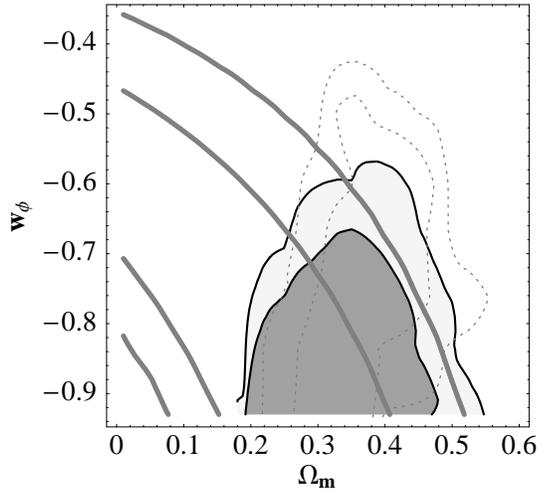}} \par}

\caption{Contour plot of the likelihood function \protect\( L(\Omega _{m},w_{\phi })\protect \).
The dotted lines are for WMAP only, the thik gray lines for the supernovae
Ia, and the gray regions are for the combined WMAP+SNIa (all contours
at the 68\% and 95\% confidence levels). }
\end{figure}

\begin{figure}[t]
{\centering \resizebox*{!}{15cm}{\includegraphics{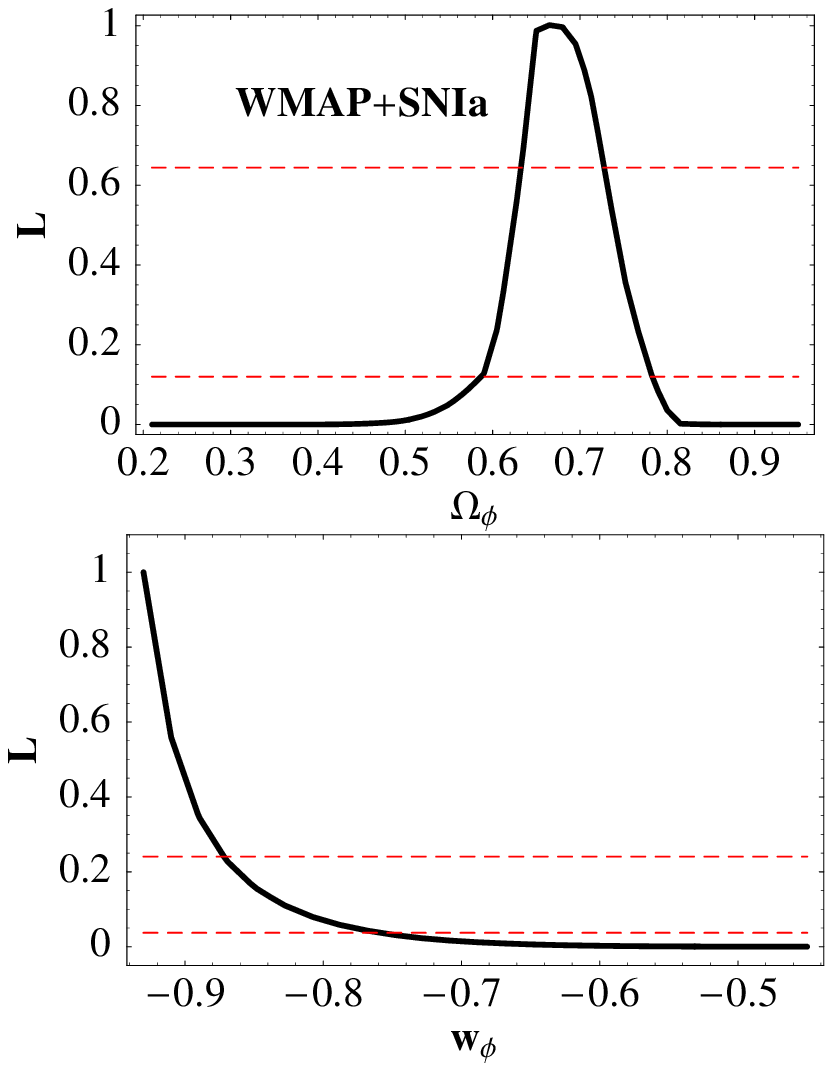}} \par}

\caption{Likelihood for \protect\( \Omega _{m}\protect \) and \protect\( w_{\phi }\protect \)
including the constraints from supernovae Ia (marginalizing over all
the other parameters, including \protect\( \beta \protect \)).}
\end{figure}

\section{Conclusions}

CMB observations are a powerful probe to constrain the properties
of dark energy. In particular, since a dark matter-dark energy interaction
would obviously escape any local gravity experiment, cosmological
observations like the CMB are the only way to observe such a phenomenon.
Since observations require the baryons to be decoupled from dark energy
(or coupled much more weakly than dark matter), the search for a non-zero
coupling \( \beta  \) is also a test of the equivalence principle.
We found that current CMB data are capable to put an interesting upper
bound to the present dark energy equation of state (all limits to
95\% c.l.)\begin{equation}
w_{\phi }<-0.49,
\end{equation}
 which becomes \( <-0.76 \) taking into account SNIa data (let us
remember that in our tracking models \( w_{\phi } \) is confined
to be \( \ge -1 \)) and to the total dark energy density,\begin{equation}
\Omega _{\phi }=0.67\pm 0.05.
\end{equation}
 For the dark matter - dark energy coupling we obtain: \begin{equation}
|\beta |<0.13,
\end{equation}
 regardless of the potential (within the tracking class we considered),
corresponding to a past equation of state \( 0\leq w_{e(\phi MDE)}<0.0075 \).
This implies that the scalar gravity is at least \( 1/\beta ^{2}\approx 60 \)
times weaker than ordinary tensor gravity. As shown in Amendola (1999),
the limit on \( \beta  \) can be restated as a limit on the constant
\( \xi  \) of the non-minimally coupled gravity, \( \xi <0.01 \).
We have shown in Amendola et al. (2003) that an experiment like the
Planck mission can lower the upper bound to \( \beta  \) to 0.05.
a limit comparable to the constraint that local gravity experiments
impose on the scalar gravity coupling to baryons, \( \beta _{baryons}^{2}<10^{-3} \)
(see e.g. Hagiwara et al. 2002).

\begin{acknowledgments}
Most of the computations have been perfomed on a 128-processors Linux
cluster at CINECA. We thank the staff at CINECA for support. We also
thanks L. Verde and H. Peiris for making available their likelihood
routine and for explanations on its use.
\end{acknowledgments}

\section*{Appendix}

Suppose we have the binned observational spectra \( C'_{\ell ,d}=(1+b_{e})C_{\ell ,d} \)
and the binned theoretical spectra \( C'_{\ell ,t}=\hat{b}_{t}C_{\ell ,t} \)
from a given experiment labelled \( e \), where \( b_{e} \) is a
calibration factor which is assumed to be distributed as a Gaussian
with zero mean and variance \( \sigma _{e} \) and \( \hat{b}_{t} \)
is the overall theoretical normalization, which is uniformly distributed
in \( (0,\infty ) \). Let \( N_{e} \) be the correlation matrix
for the \( e \)-th experiment and \begin{equation}
N'_{e}=(C_{d}^{T}C_{d})^{-1}N_{e}
\end{equation}
the approximated correlation function for the lognormal variables
\( Z\equiv \log C \) (here and in the following we suppress for clarity
the indexes \( \ell \ell'  \) of the correlation matrices and the
index \( \ell  \) from the vectors \( Z,C_{t},C_{d} \)) . If the
assumed likelihood is log-normal then we can integrate out \( b_{e} \)
as follows\begin{eqnarray}
L_{e} & = & \int \prod _{e}db_{e}e^{-\frac{1}{2}\left( \frac{b_{e}}{\sigma _{e}}\right) ^{2}}\times \\
 &  & \exp \left\{ -\frac{1}{2}\sum _{e}\left[ Z'_{t}-b_{e}-Z_{d}\right] N_{e}^{,-1}\left[ Z'_{t}-b_{e}-Z_{d}\right] \right\} =\int db_{t}\prod _{e}L_{e}
\end{eqnarray}
 where \( Z'_{t}=\log \hat{b}_{t}C_{t} \) and \( b_{e}\approx \log (b_{e}+1),Z_{d}=\log C_{d} \)
(for small \( b_{e} \)) so that \begin{equation}
L_{e}=\exp \left[ -\frac{1}{2}\left( Z'_{t}-Z_{d}\right) W_{e}^{-1}\left( Z'_{t}-Z_{d}\right) \right] 
\end{equation}
where\[
W_{e}=(C_{d}^{T}C_{d})^{-1}N_{e}+\sigma _{e}^{2}I^{T}I\]
where \( I=(1,1,1,...,n_{e}) \), \( n_{e} \) being the number of
datapoints in the \( e \)-th experiment. Note in fact that (Woodbury
formula, see e.g. Melchiorri et al. 2002)\[
(N+XX^{T})=N^{-1}-\left[ N^{-1}X(1+X^{T}N^{-1}X)^{-1}X^{T}N^{-1}\right] \]
which gives the above if \( X=\sigma _{e}I \). 

Suppose now the experimental spectra are distorted by a further \( \ell  \)-dependent
factor \( (1+w_{e}A_{\ell }) \) due to the uncertainty on the beam
angular size, where \( w \) is a random variable distributed as a
Gaussian with variance \( \sigma _{w,e} \) and \( A_{\ell } \) is
a constant vector determined by the experiment. Then each likelihood
\( L_{e} \) should be further integrated as follows\begin{eqnarray}
L'_{e} & = & \int dw_{e}L_{e}=\int dw_{e}e^{-\frac{1}{2}\left( \frac{w_{e}}{\sigma _{w,e}}\right) ^{2}}\times \\
 & = & \exp \left[ -\frac{1}{2}\left( w_{e}A+Z'_{t}-Z_{d}\right) W_{e}^{-1}\left( w_{e}A+Z'_{t}-Z_{d}\right) \right] \\
 & 
\end{eqnarray}
where, on applying again the Woodbury formula,\begin{equation}
L'_{e}=\exp \left[ -\frac{1}{2}\left( Z'_{t}-Z_{d}\right) M_{e}^{-1}\left( Z'_{t}-Z_{d}\right) \right] 
\end{equation}
where\begin{equation}
M_{e}=(C^{T}C)^{-1}N_{e}+\sigma _{e}^{2}I^{T}I+\sigma _{w,e}^{2}A^{T}A
\end{equation}
In the case of a Gaussian beam of angular size \( \theta _{0} \)
the spectrum is multiplied by a factor \( e^{\ell ^{2}\theta _{0}^{2}} \).
It can be seen then that a small misestimate \( \theta  \) (assumed
to be distributed as a Gaussian variable with variance \( \sigma _{\theta } \)
) of the beam size induces a correction on the spectrum at the \( \ell -th \)
bin by a factor \( [1+\ell ^{2}(\theta _{0}^{2}-\theta ^{2})]=1+w\ell ^{2} \),
where the random variable \( w\approx 2\theta _{0}(\theta _{0}-\theta ) \)
is also Gaussian and has uncertainty \( \sigma _{w}=2\theta _{0}\sigma _{\theta } \).
The vector \( A_{\ell } \) can be approximated then as \( \ell ^{2} \).

Finally, we integrate over \( b_{t}=\log \hat{b}_{t} \). Then we
have \( Z_{t}=b_{t}+Z'_{t} \) and\begin{eqnarray}
L=\int db_{t}\prod _{e}L'_{e}(b_{t}) & = & \int _{-\infty }^{\infty }db_{t}\exp \left[ -\frac{1}{2}\sum _{e}\left( b_{t}+Z_{t}-Z_{d}\right) M_{e}^{-1}\left( b_{t}+Z_{t}-Z_{d}\right) \right] \label{step1} \\
 & = & \exp \left[ -\frac{1}{2}\left( \gamma -\frac{\beta ^{2}}{\alpha }\right) \right] \\
 & 
\end{eqnarray}

where\begin{eqnarray*}
\alpha  & = & \sum _{e}I^{T}M_{e}^{-1}I\\
\beta  & = & \sum _{e}I^{T}M_{e}^{-1}(Z_{t}-Z_{d})\\
\gamma  & = & \sum _{e}(Z_{t}-Z_{d})^{T}M_{e}^{-1}(Z_{t}-Z_{d})
\end{eqnarray*}
which is the formula we use in this paper for the pre-WMAP data.

\end{document}